\documentstyle[editedvolume,psfig,namedreferences]{crckapb}

\begin{opening}
\title{Nonlinear \, evolution \, of \, r-modes \, in \, rotating \, 
relativistic \, stars}
\author{Jos\'e A. Font}
\institute{Max-Planck-Institut f\"ur Astrophysik\\
           Karl-Schwarzschild-Str. 1,
           85740 Garching, Germany} 
\author{Nikolaos Stergioulas}
\institute{Department of Physics, 
           Aristotle University of Thessaloniki \\
           Thessaloniki 54006, Greece}
\end{opening}

\runningtitle{Nonlinear r-modes in rotating relativistic stars}

\begin{document}

\begin{abstract}
  A numerical study of nonlinear $r$-modes in isentropic, rapidly
  rotating relativistic stars, via 3-D general-relativistic
  hydrodynamical evolutions, is presented.  On dynamical timescales,
  we find no evidence for strong coupling of $r$-modes to other modes
  at amplitudes of order one or larger. Therefore, unless nonlinear
  saturation sets in on longer timescales, the maximum $r$-mode
  amplitude is of order one. An absolute upper limit on the amplitude
  is set by causality. Our simulations also show that $r$-modes and 
  inertial modes in isentropic stars are discrete modes, with no evidence
  for the existence of a continuous part in the spectrum.
\end{abstract}

\section{Introduction}

The study of the properties of $r$-modes in rotating compact stars and
their relevance to relativistic astrophysics has received considerable
attention since the discovery (Andersson 1998, Friedman \& Morsink
1998) that these modes are unstable to the emission of gravitational
radiation.  The $r-$mode instability provides an explanation for the
spin-down of rapidly rotating young neutron stars to Crab-like
spin-periods and for the spin-distribution of millisecond pulsars and
accreting neutron stars. Additionally, it is considered to be a strong
source of continuous gravitational radiation (see Friedman \& Lockitch
1999 for a review).  Moreover, if $r$-modes induce differential
rotation, their interaction with the magnetic field in neutron stars
can enhance the toroidal magnetic field of the star (Rezzolla, Lamb \&
Shapiro 2000).

Before the instability can have an effect on the spin evolution of a
young neutron star, the $r$-mode grows to an amplitude where it is
saturated by nonlinear effects. Motivated by the absence of studies of 
such nonlinear saturation we performed a numerical study
of $r$-mode hydrodynamical evolutions in rapidly rotating relativistic 
stars. We tried to elucidate what is the maximum
amplitude such modes can reach, before nonlinear saturation
(via hydrodynamical coupling) sets in (Stergioulas \& Font 2000). The
present contribution highlights the main findings of our study. We
note that the saturation is most likely to set in on a hydrodynamical
timescale, although it cannot be excluded that weak hydrodynamical
couplings saturate the $r$-mode amplitude on longer timescales (but
shorter than the growth timescale due to gravitational radiation
reaction). However, at present, those long timescales cannot
be achieved accurately in 3-D simulations, even with the largest
available supercomputers.

\section{Numerical framework}

For our study we use a numerical code based on the 3-D \verb+CACTUS+
code (Font et al 2000, Alcubierre et al 2000), into which we
implemented the 3rd order PPM method (Colella \& Woodward 1984) for
the hydrodynamics, and initial data for equilibrium and perturbed
rapidly rotating relativistic stars (Stergioulas \& Friedman 1995). In
Font, Stergioulas \& Kokkotas (2000) it was shown that the PPM method
is suitable for long-term evolutions of rotating relativistic
stars. We focus on a representative, rapidly (and uniformly) rotating
model with gravitational mass $M=1.63M_\odot$, equatorial
circumferential radius $R=17.25$km and spin period $P=1.26$ms.  We use
the $N=1.0$ relativistic polytropic equation of state. Our simulations
employ $116^3$ Cartesian grid-points, yielding a resolution of
$0.31$km per zone.

The excitation of $r$-modes is achieved by perturbing the initial
stationary model, adding a specific 
perturbation $\delta v^i$ to the
contravariant components of the equilibrium 3-velocity, $v^i$.  As
there is no exact linear eigenfunction available in the literature
that would correspond to an $l=m=2$ $r$-mode eigenfunction for rapidly
rotating relativistic stars, we use an approximate eigenfunction,
valid in the slow-rotation $O(\Omega)$ limit to the first
post-Newtonian order (Lockitch 1999).

We also note that since $r$-modes are basically fluid modes we only
evolve the hydrodynamical variables, keeping all spacetime variables
fixed at their initial, unperturbed, values. The computational
requirements for an accurate, coupled spacetime and hydrodynamical
evolution, by far exceed current supercomputing resources.

\section{Discussion}

\begin{figure}
\centerline{ 
\psfig{file=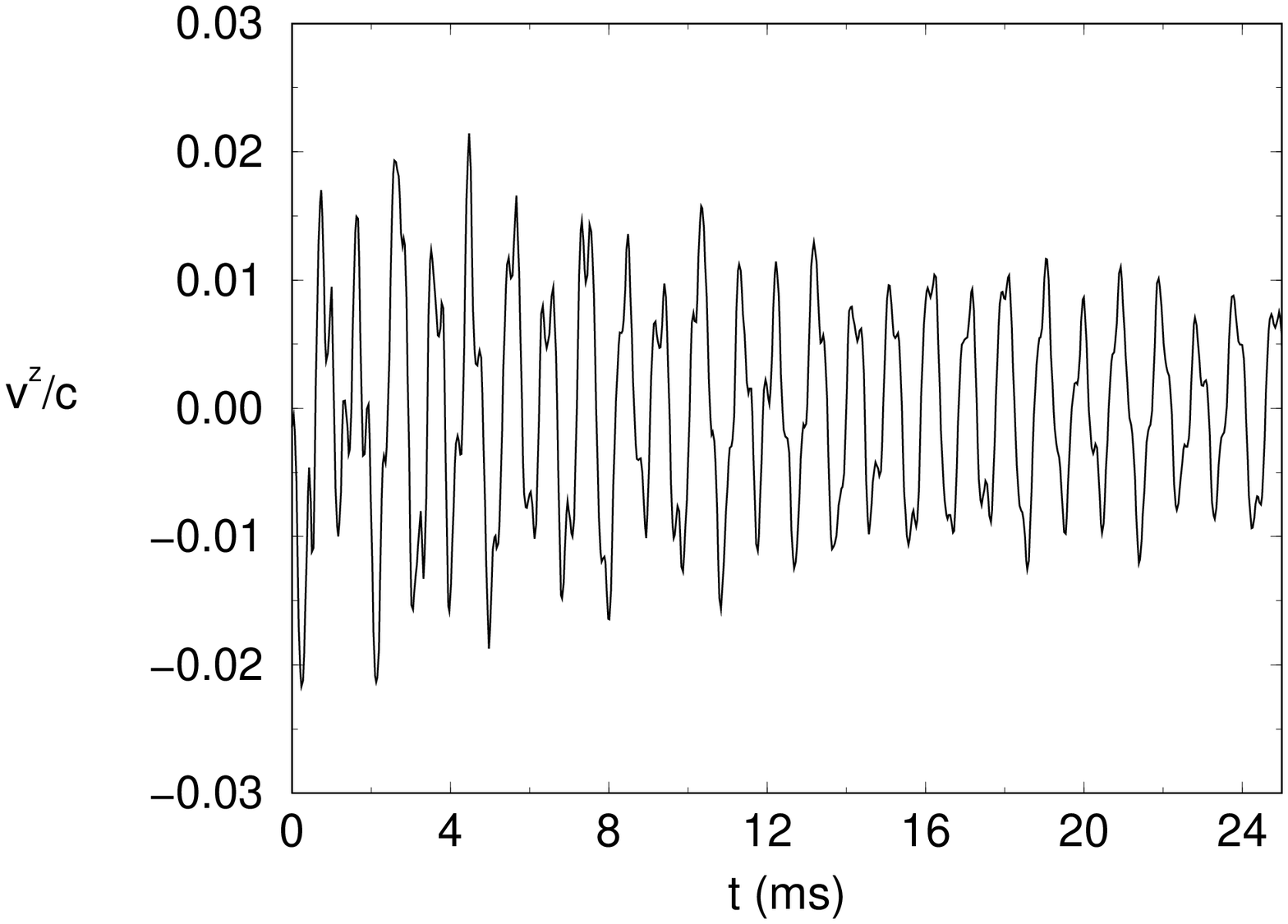,width=6.2cm,height=6.4cm}
\psfig{file=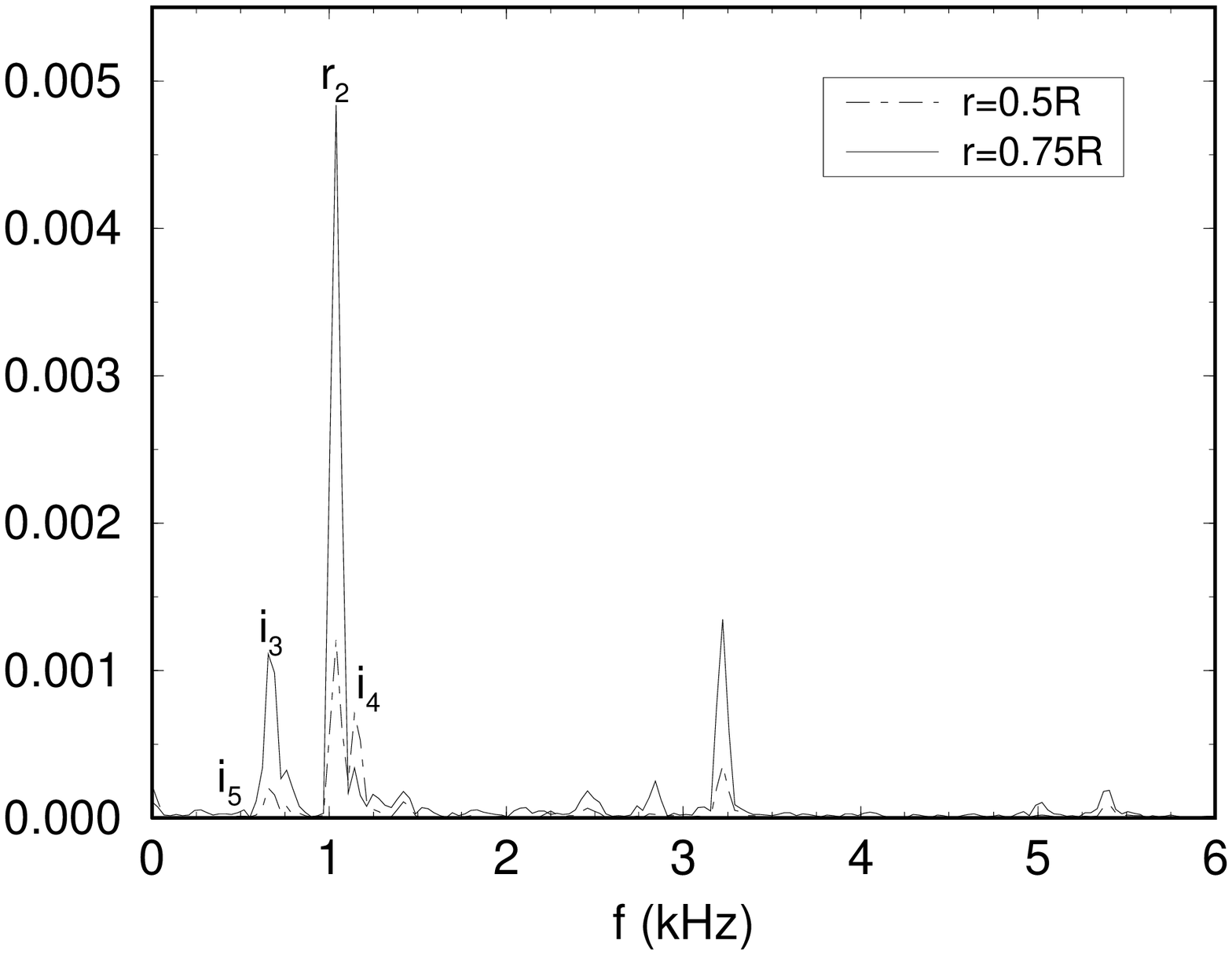,width=6.2cm,height=6.85cm}
}
\caption{{\it Left}: Evolution of the axial velocity in the equatorial 
plane for an amplitude of $\alpha=1.0$, at $r=0.75R$. The evolution is a
superposition of (mainly) the $l=m=2$ $r$-mode and several
inertial modes. The amplitude of the oscillation decreases due to
numerical (finite-differencing) viscosity of the code.
{\it Right}: Fourier transform of the velocity time-evolution,
showing the frequencies of the modes in the inertial frame. The
frequencies are the same at different radii, which
implies a discrete spectrum.}
\label{fig_2}
\end{figure}

Figure \ref{fig_2} (left panel) displays the evolution of the axial 
velocity in the
equatorial plane ($v^z$ along the $y$-axis) at a radius of $r\sim
0.75R$.  The evolution is a superposition of several modes, the
$l=m=2$ $r$-mode being the dominant component. The chosen amplitude of
the eigenfunction is $\alpha=1.0$ (in the Newtonian limit, our
definition of the amplitude agrees with that of Owen et al 1998).
The perturbed star is evolved for more than 25ms (26 $l=m=2$ $r$-mode
periods), during which the amplitude of the oscillation decreases due
to numerical viscosity only. With much larger amplitudes the evolution
is still similar to that in Figure \ref{fig_2}, with no sign of
nonlinear saturation of the $r$-mode amplitude on a dynamical
timescale. Therefore, unless nonlinear saturation sets in at
timescales much longer than the dynamical one, unstable $r$-modes
could be driven to large amplitudes (of order one) by gravitational
radiation reaction.  An absolute upper limit on the $r$-mode amplitude
is set by causality, requiring $\sqrt{v_iv^i}<c$.

A Fourier transform of the time-evolution, as a function of the
frequency in the inertial frame, is shown in the right panel of
Figure \ref{fig_2}. It
reveals that the initial data we are using, excite mainly the $l=m=2$
$r$-mode ($r_2$), with a frequency of 1.03 kHz and, with smaller
amplitudes, several inertial modes ($i_3$, $i_4$, $i_5$) and higher
harmonics. The frequencies of the various modes are the same at any
given point inside the star, which indicates that the spectrum is a
sum of discrete modes, in agreement with the conclusions of
Lockitch, Andersson \& Friedman (2000).

We have also investigated the possible appearance in our 
evolutions of the kinematical drift reported by Rezolla et al (2000).
In a rotating star with a poloidal magnetic field, this kinematical 
drift may wind up the magnetic field lines and limit the effect of 
the $r$-mode instability. Our results do indicate that the perturbed 
star is rotating slower near the surface (compared to
the unperturbed star) and that this drift scales 
roughly as $\alpha^2$ for amplitudes of order $\alpha
\sim 1.0$, although its magnitude is significantly smaller than 
that estimated by Rezzolla et al (2000).  

Our finding that gravitational radiation could drive unstable
$r$-modes to a large amplitude implies that $r$-modes can easily melt
the crust in newly-born neutron stars (Lindblom, Owen \&
Ushomirsky 1999), leaving the
initial conclusions about the $r$-mode instability being a strong
source of gravitational waves (see Friedman \& Lockitch 1999) essentially
unchanged.

\end{document}